\documentclass[%
reprint,
superscriptaddress,
altaffilletter,
showpacs,
showkeys,
amsmath, amssymb, aps, prl, floatfix,
]{revtex4-1}

\usepackage{graphicx}
\usepackage{dcolumn}
\usepackage{bm}
\usepackage{comment}
\usepackage[usenames,dvipsnames]{color}
\usepackage{hyperref}
\hypersetup{colorlinks = true, allcolors = blue}
\usepackage[mathlines]{lineno}
\usepackage{multirow}
\DeclareGraphicsExtensions{.pdf,.png,.jpg}

\newcommand{\cupidz}{CUPID-0}
\newcommand{\DBD}{0$\nu$DBD}
\newcommand{\exposure}{9.95~kg$\times$yr}
\newcommand{\exposureSe}{5.29~kg$\times$yr}
\newcommand{\exposureEmitters}{3.88$\times$10$^{25}$~$^{82}$Se nuclei$\times$yr}

\newcommand{\Tmezzi}{T$^{0\nu}_{1/2}$}
\newcommand{\background}{$(3.5^{+1.0}_{-0.9})\times10^{-3}$~counts/(keV~kg~yr)}

\begin{document}

\title{Final result of CUPID-0 phase-I in the search for the $^{82}$Se Neutrinoless Double Beta Decay}

\newcommand{\sapienza}{\affiliation{Dipartimento di Fisica, Sapienza Universit\`a di Roma, P.le Aldo Moro 2, 00185, Roma, Italy}}
\newcommand{\infnroma}{\affiliation{INFN, Sezione di Roma, P.le Aldo Moro 2, 00185, Roma, Italy}}
\newcommand{\lnl}{\affiliation{INFN  Laboratori Nazionali di Legnaro, I-35020 Legnaro (Pd) - Italy}}
\newcommand{\lngs}{\affiliation{INFN  Laboratori Nazionali del Gran Sasso, I-67100 Assergi (AQ) - Italy}}
\newcommand{\lbl}{\affiliation{Lawrence Berkeley National Laboratory , Berkeley, California 94720, USA}}
\newcommand{\infnge}{\affiliation{INFN  Sezione di Genova, I-16146 Genova - Italy}}
\newcommand{\unige}{\affiliation{Dipartimento di Fisica, Universit\`{a} di Genova, I-16146 Genova - Italy}}
\newcommand{\infnmib}{\affiliation{INFN  Sezione di Milano~-~Bicocca, I-20126 Milano - Italy}}
\newcommand{\unimib}{\affiliation{Dipartimento di Fisica, Universit\`{a} di Milano Bicocca, I-20126 Milano - Italy}}
\newcommand{\csnsm}{\affiliation{CSNSM, Univ. Paris-Sud, CNRS/IN2P3, Universit$\grave{e}$ Paris-Saclay, 91405 Orsay, France}}
\newcommand{\cea}{\affiliation{IRFU, CEA, Universit$\acute{e}$ Paris-Saclay, F-91191 Gif-sur-Yvette, France}}
\newcommand{\gssi}{\affiliation{Gran Sasso Science Institute, 67100, L'Aquila - Italy}}
\newcommand{\usc}{\affiliation{Department of Physics  and Astronomy, University of South Carolina, Columbia, SC 29208 - USA}}
\newcommand{\mpi}{\affiliation{Max-Planck-Institut für Physik, D-80805 München, Germany}}
\newcommand{\dis}{\affiliation{DISAT, Universit\`a dell'Insubria, 22100 Como, Italy}}

\author{O.~Azzolini}\lnl
\author{J.W.~Beeman}\lbl
\author{F.~Bellini}\sapienza\infnroma
\author{M.~Beretta}\unimib\infnmib
\author{M.~Biassoni}\infnmib
\author{C.~Brofferio}\unimib\infnmib
\author{C.~Bucci} \lngs
\author{S.~Capelli}\unimib\infnmib
\author{L.~Cardani}\infnroma
\author{P.~Carniti}\unimib\infnmib
\author{N.~Casali}\email[Corresponding author: ]{nicola.casali@roma1.infn.it}\infnroma
\author{D.~Chiesa}\unimib\infnmib
\author{M.~Clemenza}\unimib\infnmib
\author{O.~Cremonesi}\infnmib
\author{A.~Cruciani}\infnroma
\author{I.~Dafinei}\infnroma
\author{S.~Di~Domizio}\unige\infnge
\author{F.~Ferroni}\gssi\infnroma
\author{L.~Gironi}\unimib\infnmib
\author{A.~Giuliani}\csnsm\dis
\author{P.~Gorla} \lngs
\author{C.~Gotti}\unimib\infnmib
\author{G.~Keppel}\lnl
\author{M.~Martinez}\altaffiliation{Present address: Laboratorio de F\'isica Nuclear y Astropart\'iculas, Universidad de Zaragoza, C/ Pedro Cerbuna 12, 50009 and Fundaci\'on ARAID, Av. de Ranillas 1D, 50018 Zaragoza, Spain}\sapienza\infnroma
\author{S.~Nagorny}\altaffiliation{Present address: Queen's University, Physics Department, K7L 3N6, Kingston (ON), Canada}
\lngs\gssi
\author{M.~Nastasi}\unimib\infnmib
\author{S.~Nisi}\lngs
\author{C.~Nones}\cea
\author{D.~Orlandi}\lngs
\author{L.~Pagnanini}\unimib\infnmib
\author{M.~Pallavicini}\unige\infnge
\author{L.~Pattavina}\lngs
\author{M.~Pavan}\unimib\infnmib
\author{G.~Pessina}\infnmib
\author{V.~Pettinacci}\infnroma
\author{S.~Pirro}\lngs
\author{S.~Pozzi}\unimib\infnmib
\author{E.~Previtali}\unimib\infnmib
\author{A.~Puiu}\unimib
\author{C.~Rusconi}\lngs\usc
\author{K.~Sch\"affner}\lngs\gssi
\author{C.~Tomei}\infnroma
\author{M.~Vignati}\infnroma
\author{A.~S.~Zolotarova}\altaffiliation{Present address: CSNSM, Univ. Paris-Sud, CNRS/IN2P3, Universit$\grave{e}$ Paris-Saclay, 91405 Orsay, France}\cea

\date{\today}

\begin{abstract}
\cupidz\ is the first pilot experiment of CUPID, a next-generation project for the measurement of neutrinoless double beta decay (\DBD) with scintillating bolometers. 
The detector, consisting of 24 enriched and 2 natural ZnSe crystals, has been taking data at Laboratori Nazionali del Gran Sasso from June 2017 to December 2018, collecting a $^{82}$Se exposure of \exposureSe. 
In this paper we present the phase-I results in the search for \DBD. We demonstrate that the technology implemented by \cupidz\ allows us to reach the lowest background for calorimetric experiments: \background. Monitoring \exposureEmitters\ we reach a 90\% credible interval median sensitivity of $\rm{T}^{0\nu}_{1/2}>5.0\times10^{24}~\rm{yr}$ and set the most stringent limit on the half-life of $^{82}$Se \DBD : $\rm{T}^{0\nu}_{1/2}>3.5\times10^{24}~\rm{yr}$ (90\% credible interval), corresponding to m$_{\beta\beta} <$ (311-638)~meV depending on the nuclear matrix element calculations.

\end{abstract}


\maketitle

Nowadays, neutrinoless double beta decay (\DBD) is considered one of the most sensitive probes for Physics Beyond the Standard Model. This hypothetical nuclear transition foresees the simultaneous decay of two neutrons into protons and electrons without the emission of neutrinos \cite{Furry}. 
Its detection would prove the non-conservation of the total lepton number, setting an important milestone for leptogenesis and baryogenesis theories~\cite{Dell'Oro:2016dbc}. 
The observation of \DBD\ would give precious insights also in Particle Physics, as this transition can occur only if neutrinos coincide with their own antiparticles, according to the Majorana hypothesis. As a consequence, its detection would allow to establish the fundamental nature of these particles~\cite{Schechter:1982}. 
Furthermore, if the mechanism at the basis of \DBD\ is the exchange of light Majorana neutrinos, the half-life of the transition (\Tmezzi) would scale as T$^{0\nu}_{1/2} \propto$ m$_{\beta\beta}^{-2}$, where the parameter m$_{\beta\beta}$ (effective Majorana neutrino mass) is a superimposition of the neutrino mass eigenvalues m$_i$ weighted by the elements of the neutrino mixing matrix (U$_{ei}$): m$_{\beta\beta}$=$|\sum_{i}$U$^2_{ei}$m$_i|$~\cite{Feruglio:2002af}. Thus, a measurement of T$^{0\nu}_{1/2}$ would also allow to constrain the absolute mass scale of neutrinos.

The increasing interest in the search for \DBD\ motivated a huge international effort in the development of several technologies~\cite{Albert:2017owj,KamLAND-Zen:2016pfg,Aalseth:2017btx,Alduino:2017ehq,Agostini:2018tnm} to study some of the candidate isotopes: elements with even atomic number and even neutron number, for which the single beta decay is strongly forbidden by energy conservation law. 
These experiments are now envisioning next-generation detectors to probe half-lives exceeding 10$^{27}$~yr and, thus, the whole range of the inverted hierarchy region of neutrino masses, whose lower bound is at m$_{\beta\beta}\sim$10~meV.
The main handles to improve the sensitivity are the detector exposure, the background in the region of interest and the energy resolution~\cite{Cremonesi:2013vla}. To ensure a competitive discovery potential, next-generation detectors will need more than 10$^{27}$ emitters (hundreds of kg of source), a background as close as possible to zero in 5-10 years of data-taking, and an energy resolution of $\sim1$\%, in order to disentangle the monochromatic peak produced by \DBD\ from environmental background and from the irreducible background due to the Standard Model allowed 2$\nu$DBD decay~\cite{Alduino:2017qet}.

Cryogenic calorimeters (historically also called bolometers~\cite{Fiorini:1983yj}) are one of the leading technologies in the search for \DBD. 
These devices consist of dielectric and diamagnetic crystals cooled down at cryogenic temperatures (10~mK) and equipped with a sensor to convert thermal variations into electrical signals. The sensors chosen for CUPID-0 are Neutron Transmutation Doped (NTD) Ge thermistors~\cite{thermistor}.
The most impressive implementation of this technology is CUORE~\cite{Artusa:2014lgv}, an array of 988 TeO$_2$ cryogenic calorimeters, for a total active mass of 742~kg, that is taking data at Laboratori Nazionali del Gran Sasso (LNGS, Italy) to study the $^{130}$Te \DBD. With an average energy resolution of 7.7$\pm$0.5~keV in the region of interest, in 2018 CUORE set a 90\% credible interval (C.~I.) lower limit on \Tmezzi\ of 1.5$\times$10$^{25}$~yr~\cite{Alduino:2017ehq}. 
The success of CUORE motivated the proposal of CUPID (CUORE Upgrade with Particle IDentification~\cite{Wang:2015raa,Wang:2015taa,Artusa:2014wnl,Alduino:2016vtd}), which aims at exploring the inverted hierarchy region of neutrino masses by reaching a \DBD\ sensitivity larger than 10$^{27}$~yr. CUPID will increase the active mass through isotopic enrichment and, at the same time, decrease the background by two orders of magnitude with respect to CUORE. Since the CUORE background in the region of \DBD\ is dominated by $\alpha$ decays occurring on the detector surfaces~\cite{Alduino:2017qet}, the first milestone of CUPID is the development of an active technique which enables the identification of such interactions. 
In the last years, scintillating bolometers proved to be a viable path for the $\alpha$ background suppression, as the simultaneous measurement of the calorimetric signal and the scintillation light allows to keep a high energy resolution while identifying the nature of the interaction~\cite{Bobin:1997qm,Pirro:2005ar}. 
The promising R$\&$D activity performed by the LUCIFER~\cite{Beeman:2013sba,Beeman:2013vda,Beeman:2012ci,Beeman:2012jd,Beeman:2011bg,Cardani:2013mja,Cardani:2013dia,Artusa:2016maw} and LUMINEU~\cite{Barabash:2014una,Armengaud:2017hit,Buse:2018nzg,Armengaud:2015hda,Bekker:2014tfa} experiments in the development of scintillating bolometers allowed to construct a first medium-scale demonstrator of this technology: \cupidz.

The \cupidz\ collaboration focused on $^{82}$Se, which features one of the highest Q-values for \DBD\ decay (2997.9$\pm$0.3 keV~\cite{Lincoln:2012fq}). This choice allows to work in an energy region barely affected by the $\beta/\gamma$ background induced by the natural radioactivity, that drops of more than an order of magnitude above 2.6~MeV.
The main limitation of $^{82}$Se, i.e. its poor natural isotopic abundance, can be overcome through isotopic enrichment. The \cupidz\ collaboration exploits 96.3$\%$ enriched Se to synthesise ZnSe powder~\cite{Beeman:2015xjv} 
 and grow 24 cylindrical Zn$^{82}$Se crystals 95\% enriched in $^{82}$Se~\cite{Dafinei:2017xpc}. 
These crystals (plus two natural ones) are surrounded by 3M Vikuiti plastic reflective foils and interleaved by bolometric light detectors (LD) similar to those described in Ref.~\cite{Beeman:2013zva}. Both the ZnSe crystals and the LD are equipped with a NTD Ge thermistor and with a P-doped Si Joule heater injecting a periodic reference pulse to correct for temperature variations~\cite{Arnaboldi:2003yp,Andreotti2012}.
The detectors are arranged in five towers using a mechanical copper structure, and anchored to the coldest point of an Oxford $^{3}$He/$^{4}$He dilution refrigerator located in the Hall-A of LNGS (previously used for the CUORE-0 experiment~\cite{Alfonso:2015wka}). Details about the detector construction, commissioning and operation can be found in Ref.~\cite{Azzolini:2018tum}.

The \cupidz\ detector has been taking data from June 2017 to December 2018. Physics data are divided in nine blocks called ``data sets''. At the beginning and at the end of each data set we perform an energy calibration exposing the detector to an external (removable) $^{232}$Th source which provide $\gamma$ lines up to 2615~keV. For the \DBD\ search, we discard from the analysis the data acquired by the two natural crystals and also by two enriched crystals that are not featuring a satisfactory bolometric performance. Thus, the final active mass used for the described analysis is 8.74~kg, corresponding to 4.65~kg of $^{82}$Se. With a live-time for physics runs of 74\%, we collected a total $^{82}$Se exposure of \exposureSe\ (\exposureEmitters). The first four data sets, referring to 1.83~kg$\times$yr of $^{82}$Se, were used to extract a preliminary limit on the half-life of $^{82}$Se \DBD\ both to the ground and exited states of $^{82}$Kr~\cite{Azzolini:2018dyb,Azzolini:2018oph}. Since then, we increased by almost a factor 3 the statistics and we improved the analysis algorithms to enhance the energy resolution and the background discrimination capability. 
In this paper we present the final result of \cupidz\ phase-I for the search of \DBD.

The signals produced by particle interactions are amplified, filtered with a 120 dB/decade, six-pole anti-aliasing active Bessel filter, and fed in a 18 bit analog-to-digital converter board with a sampling frequency of 1~ksps for ZnSe and 2~ksps for the (faster) LDs. The electronics is described in Refs.~\cite{Arnaboldi:2018yp,Carniti2016,arnaboldi20018,arnaboldi2015,arnaboldi2010,Arnaboldi:2004jj,AProgFE}. Data are saved on disk using a custom DAQ software package~\cite{DiDomizio:2018ldc}. The ZnSe signals are acquired using a derivative trigger with channel-dependent parameters. Since the pulses recorded by the LD have a lower signal to noise ratio (as the majority of the energy is released in the heat channel), we used a different trigger, forcing the acquisition of the LD waveforms every time the trigger of the corresponding ZnSe fires.

The data analysis follows a slightly different path for heat and light pulses. 
Heat pulses recorded by the ZnSe are processed using a software matched filter algorithm~\cite{Gatti:1986cw,Radeka:1966}, corrected by temperature instabilities using the reference pulse periodically injected by the Si resistor, and energy-calibrated. To this aim, we exploit the periodical calibration with $^{232}$Th sources: we identify the most intense peaks produced by the sources and use a zero-intercept parabolic function to convert the signal amplitude into energy. Finally, we compute time-coincidences among crystals using a coincidence window of 20~ms, in order to reject the background due to multi-site events simultaneously triggering different detectors.

Light pulses are acquired in time-coincidence with ZnSe and filtered using a slightly modified version of the matched filter algorithm~\cite{Azzolini:2018yye}. After improving the amplitude estimation by taking advantage from the known time jitter between heat and light channels~\cite{Azzolini:2018yye}, we inter-calibrate the light detectors using the 2.6~MeV line produced by the $^{232}$Th sources. Finally, we remove the correlation between light and heat signals by performing a rotation in the light energy vs. heat energy scatter plot to enhance the energy resolution~\cite{Beretta:2019bmm}.

\begin{figure}[tb]
\centerline{\includegraphics[width=8.5cm]{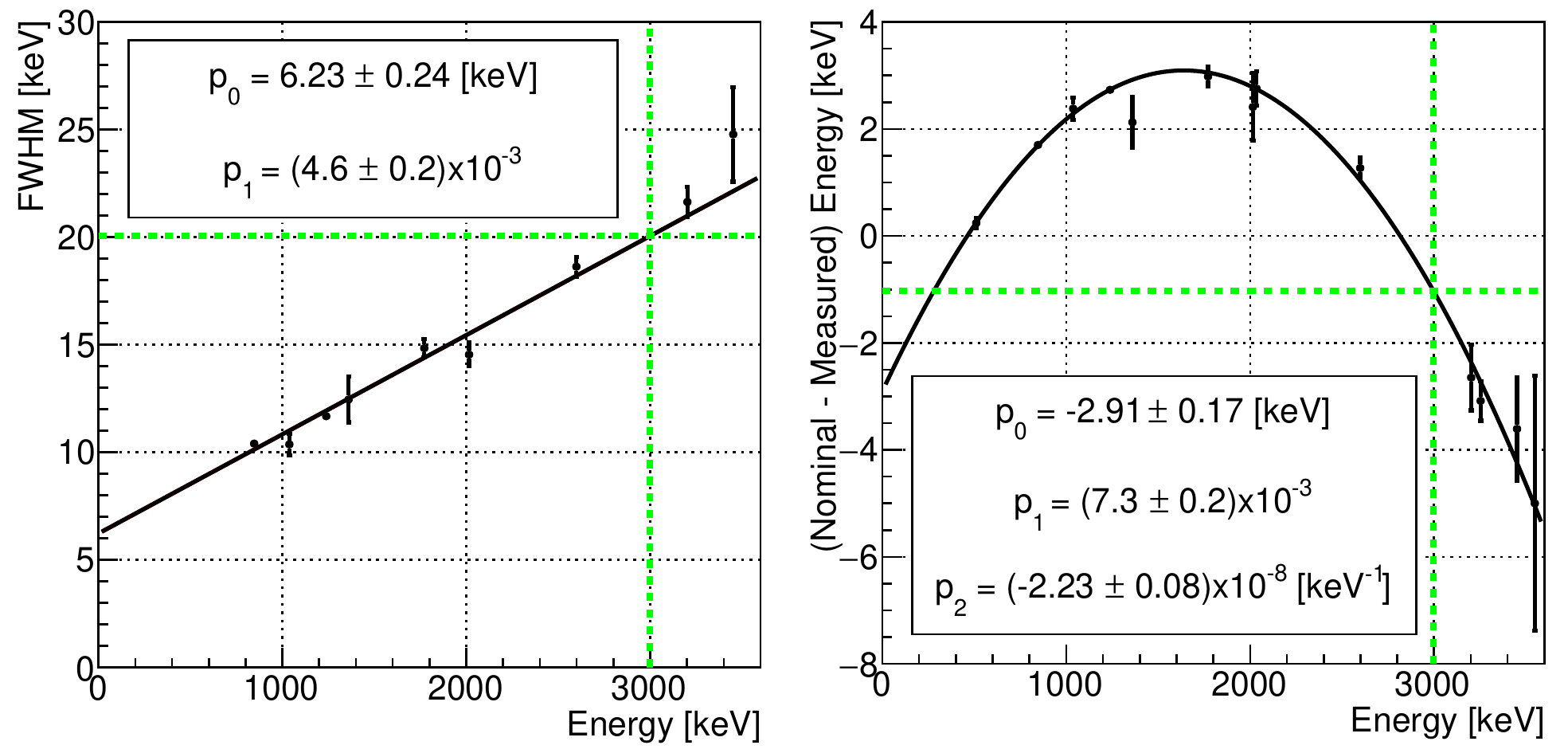}}
\caption{Left: energy resolution as a function of the energy for the most intense $\gamma$ peaks emitted by a $^{56}$Co source (sum spectrum of all the ZnSe channels). Right: residuals of the calibration made with the coefficients derived from the $^{232}$Th calibrations. Green cross: values corresponding to the $^{82}$Se Q-value, where we expect an energy resolution of 20.05$\pm$0.34~keV and a residual of the calibration of -1.03$\pm$0.18~keV.}
\label{fig:calibration}
\end{figure}

To check the goodness of the energy reconstruction we made a calibration with a $^{56}$Co source (T$_{1/2}\sim77.3$ days, Q-value~$\sim$~4.6~MeV) emitting $\gamma$ rays with significant branching ratio up to 3.3~MeV. We calibrate the spectrum of each crystal using the coefficients derived from the $^{232}$Th calibrations, then apply these calibration coefficients to the runs performed with the $^{56}$Co source and obtain the sum spectrum of all the calibrated channels. 
To derive the FWHM energy resolution and the residuals of the calibration, we fit the most prominent $\gamma$ peaks of the spectrum. Bolometric detectors often show a non perfectly Gaussian response to monochromatic energy deposits~\cite{Alduino:2016zrl,Casali:2013zzr}. For this reason, instead of using a single Gaussian function, we fit the $^{56}$Co peaks using the signal model of CUPID-0 that, as explained in Ref~\cite{Azzolini:2018yye}, consists in a double Gaussian function. The FWHM energy resolution and the residuals of the calibration are shown in Fig.~\ref{fig:calibration}.

The energy resolution is modelled as a function of the energy using a linear dependency. The value extracted at the Q-value of $^{82}$Se is 20.05$\pm$0.34~keV FWHM, about 10$\%$ better with respect to the one used for the analysis described in Ref.~\cite{Azzolini:2018dyb} thanks to the de-correlation of heat and light signals~\cite{Beretta:2019bmm}.

The residuals of the $^{56}$Co peaks show a small parabolic dependency on the energy, that could be removed by a further improvement of the calibration function. Nevertheless, the residual at the Q-value of $^{82}$Se is $-1.03\pm0.18$~keV, which is negligible considering the $\sim$20 keV energy resolution. For this reason, instead of acting on the calibration function to further decrease the residual at the Q-value, we decided to treat this parameter as a systematic uncertainty. 

We monitored possible drifts in the calibration parameters during all the physics runs with periodical $^{232}$Th calibrations, and we observed variations in the peaks positions lower than $\pm$0.1$\%$. The use of the $^{56}$Co calibration allowed us to narrow the systematic uncertainty on the \DBD\ position by a factor 3 compared to the previous analysis described in Ref.~\cite{Azzolini:2018dyb}. Nonetheless, this technique could not be used as routine calibration for CUPID-0 
for the sake of physics data live time, since the time needed to collect sufficient $\gamma$ peaks near our region of interest was of the order of three weeks.

To select the events of interest for the \DBD\ search, we apply the following data selection criteria. First, we reject non-physics events (due for example to electronics noise) by applying basic cuts to the shape parameters of the heat pulses, such as the slope of the baseline before the time of interaction, and the rise-time of the pulse. We also require that a single crystal in the array triggered that event, as from GEANT4 simulations we expect the two electrons of \DBD\  to be contained in a single ZnSe crystals with a probability of (81.0$\pm$0.2)$\%$

Then, we select electron-like events by exploiting the time-development of the light pulses which, being dominated by the crystal scintillation, allows to discriminate electrons from $\alpha$ particles. To this purpose we use a shape parameter defined in Ref.~\cite{Azzolini:2018yye}, which is very sensitive to the time development of the light pulse. To improve the discrimination capability we combine the shape parameters of the two light detectors: since they result to be uncorrelated, we simply average them. To select a pure $\beta/\gamma$ sample, we use electromagnetic showers induced by muons interacting in the material close to the detector, easily recognisable as they induce simultaneous triggers in more than one detector. In order to make this selection more robust, we only choose events triggered by more than four crystals. 
We set a cut on the shape parameter of light pulses in order to keep a signal efficiency of 98$\%$. The probability for an $\alpha$ particle to pass this selection is lower than 10$^{-7}$, proving the potential of the light and heat read-out in suppressing the $\alpha$ background in view of the next-generation experiment CUPID.
After the removal of  $\alpha$ particles, the background in the region of interest decreases from 3.2$\times$10$^{-2}$~counts/(keV~kg~yr) to 1.3$\times$10$^{-2}$~counts/(keV~kg~yr). 
The residual background is dominated by $^{208}$Tl decays due to internal and surface contaminations of the crystals.

In CUPID-0, this background can be suppressed with a final data selection, made by applying a time-veto.
Indeed, if the contamination is internal or very close to the crystal, it is possible to tag the mother of $^{208}$Tl ($^{212}$Bi) thanks to its $\alpha$ decay, and then exploit the short half-life of $^{208}$Tl (3.05~min) to reject subsequent $\beta$ and $\gamma$ deposits. If the contamination is internal, the decay of $^{212}$Bi is expected to produce an $\alpha$ event with an energy corresponding to the Q-value of the transition; if the contamination is on the crystal surface, part of the energy can be lost, and a $^{212}$Bi decay results in an $\alpha$ event with lower energy. In \cupidz\ we can exploit the particle identification capability to tag all the $\alpha$ particles down to 2~MeV, enhancing the rejection of $^{212}$Bi - $^{208}$Tl events. 

Then we open a time-veto of 7 $^{208}$Tl half-lives for each tagged $\alpha$ event. When an $\alpha$ particle is detected, the probability for a $^{208}$Tl decay to occur after 7 half-lives is 0.8$\%$. Thus, a time-veto of 7 half-lives offers a tagging efficiency exceeding 99\% with a reasonable dead-time of 6$\%$. We highlight that, despite the strong particle identification capability offered by \cupidz, there is still a small probability that surface $\alpha$ particles escape detection. According to our background model, surface contaminations of the crystals in Th produce a residual background at the level of 3.4$\times$10$^{-4}$~counts/(keV~kg~yr)~\cite{Azzolini:2019nmi}.

\begin{figure}[htb]
\centerline{\includegraphics[width=8.5cm]{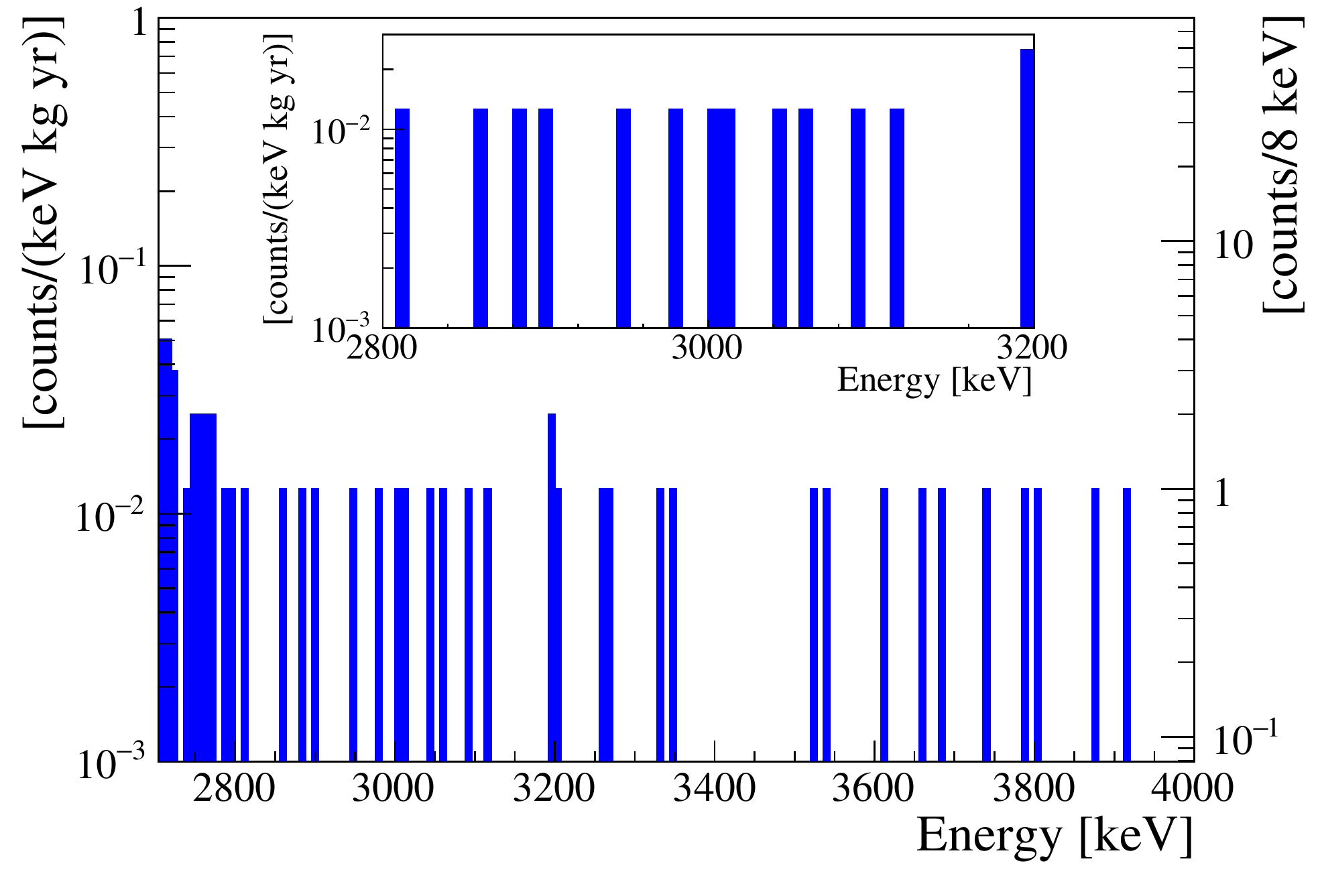}}
\caption{Background measured by CUPID-0 (exposure of \exposure\ of Zn$^{82}$Se crystals) after the three selection steps described in the text. The events below 2.8 MeV come from the tail of $2\nu\beta\beta$ decay. Inset: energy region of interest for the \DBD\ search: from 2.8 to 3.2 MeV, a symmetric energy region around the Q-value of $^{82}$Se (2997.9$\pm$0.3~keV).}
\label{fig:background}
\end{figure}

The total efficiency on the signal selection thus includes:
\begin{itemize}
\item the trigger and energy reconstruction efficiency ($\sim99.5$\%), computed using the reference pulses injected by the Si resistor (as done, for example, in Ref.~\cite{Alduino:2016zrl});
\item the data selection efficiency of heat pulses, comprising the efficiency of the cuts on the pulse shape, and the dead-time introduced by the time-veto of 7 half-lives used to reject $^{212}$Bi - $^{208}$Tl events ($\sim88$\%). This number was evaluated using the $\sim$1~MeV $\gamma$-peak produced by $^{65}$Zn decay in the physics spectrum~\cite{Azzolini:2018yye};
\item the $\beta/\gamma$ selection efficiency ($\sim98\%$), computed on a sample of pure $\gamma$ events. To define this event class, we exploit electromagnetic showers induced by cosmic muons; such events are easily identifiable as they result in interactions in a large number of crystals (see details in Ref.~\cite{Azzolini:2018yye}).
\end{itemize}
The total efficiency, averaging the nine data sets, summed up to (86$\pm$1)$\%$. This number is combined with the (81.0$\pm$0.2)$\%$ probability that the two electrons emitted in \DBD\ are contained in a single crystal. Thus, the final \cupidz\ signal efficiency results (70$\pm$1)\%.
The background spectrum between 2.7 MeV and 4.0 MeV after the three selection steps is shown in Fig.~\ref{fig:background}.

In the ROI for the \DBD\ search (from 2.8 to 3.2~MeV, see inset in Fig.~\ref{fig:background}) we observe 14 counts. Performing a simultaneous unbinned extended maximum likelihood (UEML) fit in the ROI, modelled with the only contribution of a flat background, we obtain \background, the lowest background achieved by \DBD\ experiments based on cryogenic calorimeters. We then repeat the simultaneous UEML fit in the ROI adding the signal hypothesis to the model. For each data set, the fit includes the bi-Gaussian line shape for the \DBD\ signal~\cite{Azzolini:2018dyb}, and the flat background component. The efficiency, the exposure and the energy resolution are data set dependent. On the contrary, the decay rate $\Gamma^{0\nu}$ ($\Gamma^{0\nu}= 1/T^{0\nu}_{1/2}$ ) and the background index are treated as free parameters common to all the detectors and data sets. 

\begin{figure}[htb]
\centerline{\includegraphics[width=8.5cm]{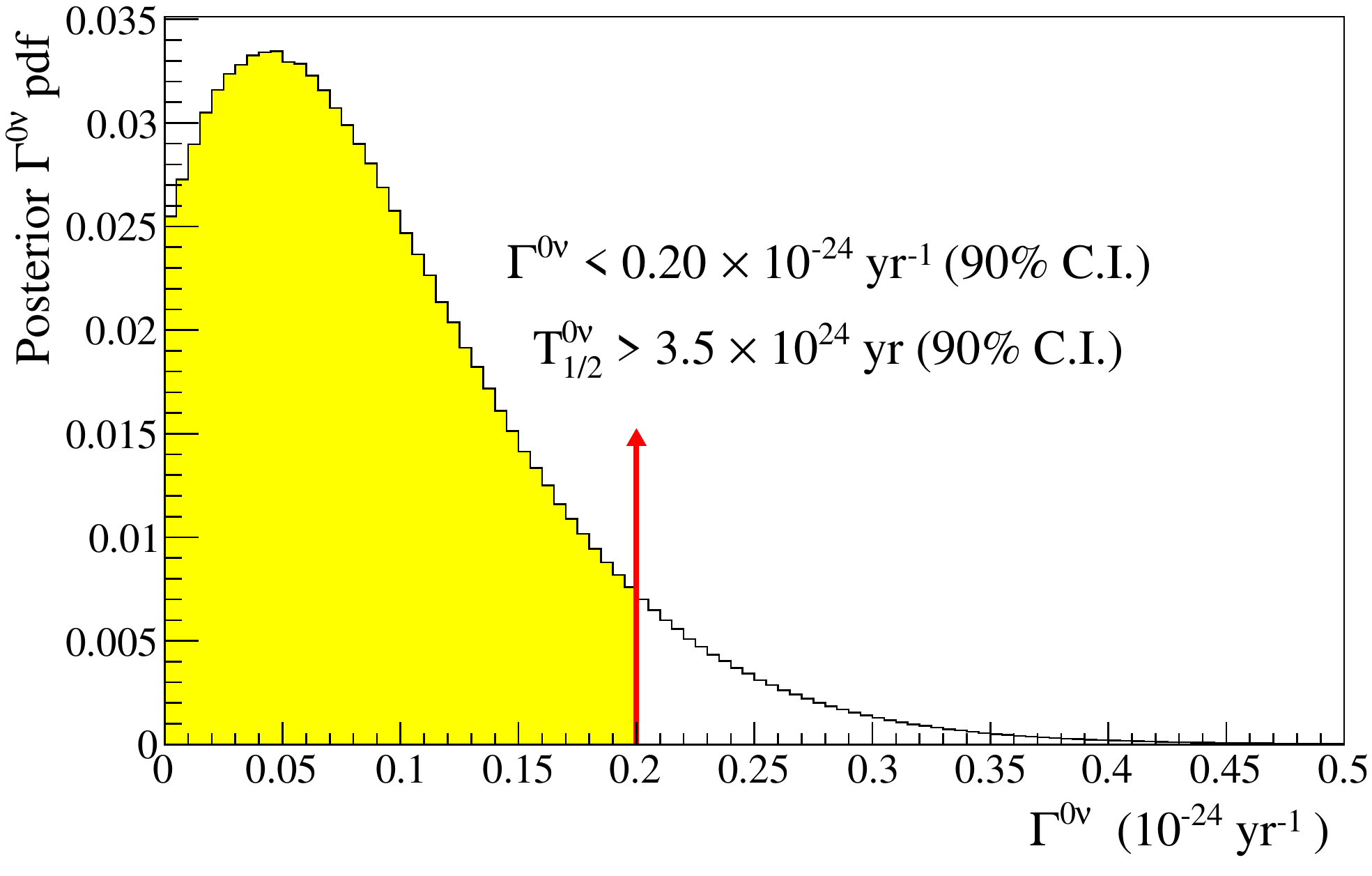}}
\caption{Posterior probability density function (pdf) for $\Gamma^{0\nu}$ as results from the simultaneous UEML fit in the ROI after the marginalization over the background index nuisance parameter. The posterior pdf accounts for all the statistics and systematic uncertainties arising from the energy scale, detector resolutions, efficiencies and exposures measured for each data set. The yellow histogram represents the 90\% area of the posterior pdf.}
\label{fig:posteriorPDF}
\end{figure}

We account for the systematics due to the uncertainty on the energy scale, the detector resolutions, the efficiencies, and the exposures. For each influence parameter, we weight the likelihood with a Gaussian probability density function with the mean and width fixed to the best estimated values and uncertainties, respectively. 
In contrast to the analysis described in Ref.~\cite{Azzolini:2018dyb}, we treat the energy resolution and the uncertainty on the energy scale separately, as the former is a data-set dependent parameter, while the latter refers to the entire statistics. 
We then integrate the likelihood via numerical integration using a uniform prior in the physical region of $\Gamma^{0\nu}$ (between 0 and 0.5~$\times10^{-24}$~yr$^{-1}$) and marginalize over the background index nuisance parameter. The resulting posterior probability density function (pdf) is shown in Fig.~\ref{fig:posteriorPDF}. The fit converges to a value of $\Gamma^{0\nu}$ of $4.7^{+7.9}_{-5.3}\times10^{-26}$~yr$^{-1}$. We found no evidence for \DBD\ and we set a 90\%~C.~I. Bayesian upper limit on $\Gamma^{0\nu}<0.20\times10^{-24}$~yr$^{-1}$, corresponding to a lower limit on the $^{82}$Se half-life of
\begin{equation}
\rm{T}^{0\nu}_{1/2}>3.5\times10^{24}~\rm{yr}.
\end{equation}
We also take into account the hypothesis of an exponential background distribution instead of a flat one, obtaining the same results.

Exploiting a toy MC in which we simulate 1000 experiments with the same performances and exposure measured by CUPID-0, we evaluate the 90$\%$~C.~I. median sensitivity to be $\rm{T}^{0\nu}_{1/2}>5.0\times10^{24}~\rm{yr}$. We evaluate also the probability to reach a higher limit with respect to the one measured in this work, obtaining 80\%. 
As far as the limit on the effective neutrino mass is concerned, we recall that in the models where the \DBD\ is mediated by the exchange of a light Majorana neutrino, the effective neutrino mass m$_{\beta\beta}$ is related to T$^{0\nu}_{1/2}$ by:
\begin{equation}
   \mathrm{(T^{0\nu}_{1/2})^{-1}  = G_{0\nu} ~ |\mathcal{M}_{0\nu}|^2 ~ \left(g_{a}\right)^4 ~ \left(\frac{m_{\beta\beta}}{m_e}\right)^2}
   \label{eq:t12tombb}
\end{equation}
where G$_{0\nu}$, M$_{0\nu}$, g$_{\mathrm{a}}$ and m$_{\mathrm{e}}$ are the phase space factor of the decay, the dimensionless nuclear matrix element (NME), the effective axial coupling constant and the electron mass respectively. Using  G$_{0\nu}$  from Refs.~\cite{Kotila:2012zza}, the NME from Refs.~\cite{Engel:2016xgb,Yao:2014uta,Menendez:2008jp,Simkovic:2013qiy,Rodriguez:2010mn,Meroni:2012qf} and an axial coupling constant g$_{\mathrm{a}}$=1.269 we set an upper limit on m$_{\beta\beta}<~$(311-638)~meV.

Summarizing, \cupidz\ set the most stringent limit on the \DBD\ half-life of $^{82}$Se. Thanks to the capability to reject the background coming from $\alpha$ radioactive contaminations granted by the scintillating cryogenic calorimeter technique, we reach an unprecedented background level for bolometric experiments: \background.

According to our background reconstruction model~\cite{Azzolini:2019nmi}, the main contribution in the ROI is due to $\mu$ interactions: $(1.53\pm0.13~\rm{stat}\pm0.25~\rm{syst})\times10^{-3}$~counts/(keV~kg~yr). The contaminations of the crystals dominate the residual background. In order to assess the individual contributions to the measured background, at the beginning of 2019 we upgraded the \cupidz\ detector by installing a muon veto and removing the reflecting foil that prevents the analysis of coincidences of surface events among crystals. In such a way, we plan to improve the capability of recognize the sources of the $\beta/\gamma$ background measured by \cupidz. Measuring residual background contributions with such a high sensitivity will be of crucial importance in anticipation of the next-generation CUPID experiment.

This work was partially supported by the European Research Council (FP7/2007-2013), Contract No. 247115. We are particularly grateful to S. Grigioni and M. Veronelli for their help in the design and construction of the sensor-to-absorber gluing system, M. Iannone for the help in all the stages of the detector construction, A. Pelosi for the construction of the assembly line, M. Guetti for the assistance in the cryogenic operations, R. Gaigher for the calibration system mechanics, M. Lindozzi for the development of cryostat monitoring system, M. Perego for his invaluable help,  the mechanical workshop of LNGS (E. Tatananni, A. Rotilio, A. Corsi, and B. Romualdi) for the continuous help in the overall setup design. We acknowledge the Dark Side Collaboration for the use of the low-radon clean room.
This work makes use of the DIANA data analysis and APOLLO data acquisition software which has been developed by the CUORICINO, CUORE, LUCIFER, and CUPID-0 Collaborations.

\end{document}